\documentclass[11pt]{article}

\usepackage{color}		
\usepackage[dvips]{graphicx}	
\usepackage{amsfonts}		
\usepackage{amsmath}		
\usepackage{amssymb}		
\usepackage{wasysym}
\usepackage{multirow}
\usepackage{mciteplus}
\usepackage{psfrag}
\usepackage{url}
\usepackage{soul}
\usepackage[normalem]{ulem}

\include{paperdef}

\newcommand*{\figref}[1]{Fig.~\ref{#1}}

\begin{document}
\hfill {\tt DESY 11-243}

\def\thefootnote{\fnsymbol{footnote}}

\begin{center}
\Large\bf\boldmath
\vspace*{2cm} Interpreting the LHC Higgs Search Results in the MSSM
\unboldmath
\end{center}
\vspace{0.4cm}
\begin{center}
S.~Heinemeyer$^{1,}$\footnote{Electronic address: Sven.Heinemeyer@cern.ch},
O.~St{\aa}l$^{2,}$\footnote{Electronic address: oscar.stal@desy.de},
G.~Weiglein$^{2,}$\footnote{Electronic address: Georg.Weiglein@desy.de} \\[0.4cm] 
\vspace{0.4cm}
{\sl$^1$Instituto de F\'isica de Cantabria (CSIC-UC), Santander,  Spain
}\\[0.2cm]
{\sl $^2$ Deutsches Elektronen-Synchrotron DESY\\
 Notkestra{\ss}e 85, D-22607 Hamburg, Germany}\\[0.2cm]
\end{center}
\vspace{0.2cm}

\renewcommand{\thefootnote}{\arabic{footnote}}
\setcounter{footnote}{0}

\begin{abstract}
Recent results reported by the ATLAS and CMS experiments on the
search for a SM-like Higgs boson both show an excess for a Higgs mass
near $125\gev$, which is mainly driven 
by the $\gamma\gamma$ and $ZZ^*$ decay channels, but also receives some support
from channels with a lower mass resolution.
We discuss the implications of this possible signal within
the context of the minimal supersymmetric
Standard Model (MSSM), taking into account previous limits from Higgs
searches at LEP, the Tevatron and the LHC. The consequences for the
remaining MSSM parameter space are investigated. Under the assumption of a Higgs signal we derive new lower bounds on the tree-level parameters of the MSSM Higgs sector. We also discuss briefly an alternative interpretation of the excess in terms of the heavy CP-even Higgs boson, a scenario which is found to be still viable.
\end{abstract}
\newpage


\section{Introduction}
The Higgs boson \cite{Higgs:1964pj,*Englert:1964et,*Higgs:1964ia} has for a long time been considered as the only missing piece in the Standard Model (SM) of particle physics. Therefore, finding this particle has been one of the main tasks of experimental
high-energy physics.  However, the main results from the 
published searches so far have been exclusion limits
(see e.g.~the results from LEP \cite{Barate:2003sz}, the Tevatron \cite{TevHiggs}, and the LHC \cite{CMSHiggs,ATLASHiggs}). Combining the experimental limits, the only
allowed region (before the latest results which will be discussed below) a relatively small window for the Higgs mass: 
$114 \gev < \MHSM < 141 \gev$. 
This low mass region is also the one favoured by electroweak precision tests, see
e.g.~\cite{lepewwg,*Gfitter}.  

A low Higgs mass is predicted in supersymmetric extensions of the
SM, where the quartic Higgs couplings are related to gauge
couplings. Exclusion of a heavy SM-like Higgs~\cite{TevHiggs,CMSHiggs,ATLASHiggs} can
therefore be considered as being in line with the predictions
of supersymmetry (SUSY). Besides predicting a light Higgs boson, SUSY protects scalar masses from
the large hierarchy of scales, it allows for gauge coupling unification,
and it can provide a dark matter candidate~\cite{Goldberg:1983nd,*Ellis:1983ew}. The \emph{minimal}
supersymmetric extension of the SM (MSSM)~\cite{Nilles:1983ge,*Haber:1984rc,*Barbieri:1987xf} has two complex Higgs
doublets. Following electroweak symmetry breaking, the physical spectrum
therefore contains five Higgs bosons. Assuming CP conservation, these
are denoted $h,H$ (CP-even), $A$ (CP-odd), and $H^\pm$ (charged
Higgs). At the tree-level the MSSM Higgs sector can be described by two parameters (besides the SM
parameters), commonly chosen as the mass of the CP-odd Higgs boson, $\MA$, and $\tb$, the
ratio of the two vacuum expectations values. 
In the decoupling limit, $\MA \gsim 2 \MZ$ (where $\MZ$ denotes the mass
of the $Z$~boson), all MSSM Higgs bosons except the lightest
CP-even scalar $h$ become heavy, whereas $h$ has SM-like
properties. In this limit it would be difficult to separate hints for a SM
Higgs boson from a potential MSSM counterpart. It is also
in the decoupling limit where $\Mh$ reaches its maximal value, 
$\Mh\simeq 135\gev$~\cite{Degrassi:2002fi}.

The LHC experiments recently extended their exclusion
regions for a SM-like Higgs boson down to $\MHSM \lesssim 127 \gev$, 
with the
lowest limit coming from CMS ($\MHSM < 131 \gev$ for ATLAS). 
In addition, ATLAS
reported exclusion of the range $114 \gev < \MHSM < 115.5 \gev$, 
which is a region where sensitivity was not expected. Most interestingly, both experiments 
also reported about an excess over the background expectation close to
$\MHSM = 125\gev$ \cite{Dec13}. Since this Higgs mass lies in the range compatible
with supersymmetry, we report in this letter on a first analysis and
interpretation of these results in an MSSM context.


\section{Experimental Higgs search results}
Both the LHC experiments (ATLAS and CMS) have reported \cite{Dec13} on indications for an
excess of Higgs-like events corresponding to a Higgs boson mass%
\footnote{Another excess at $\MHSM\simeq 119 \gev$ was reported by CMS, but not
  confirmed by ATLAS. Consequently, we will not consider this value in
  our analysis.}
\begin{align*}
\MHSM &= 126\gev\qquad\mathrm{(ATLAS)},\\
\MHSM &= 124\gev\qquad\mathrm{(CMS)}.
\end{align*}
The result is driven by an observed excess of events over SM background
expectations in primarily the $\gamma\gamma$ and $ZZ^*$ channels, which provide relatively good resolution for the Higgs boson mass. 
The local significance for the combined result is $3.6\,\sigma$ for ATLAS and
$2.6\,\sigma$ for CMS. However, when interpreted in a global search
containing many mass bins, the local significance is washed out by the
look-elsewhere effect (LEE). This effect compensates for the higher
probability of random fluctuations generating an excess anywhere when
searching in more than one place. Taking this into account, the
significance of the reported result is reduced to $2.5\,\sigma$
($1.9\,\sigma$) for ATLAS (CMS) when interpreted as a SM Higgs 
search over the mass range from $110\gev$ to $146\gev$. 
On the other hand, one could argue that when interpreting
these results in a model where the allowed range for $\Mh$ is constrained to a
smaller range by the theory (as in the MSSM), the LEE does not apply to the
same degree as for the SM interpretation. These new results are therefore even
somewhat more interesting in an MSSM context.  

For the remainder of this paper, encouraged by the excess reported by
ATLAS and CMS, we investigate a scenario where we assume the observation
of a state compatible with a SM-like Higgs boson with mass 
$\Mh=(125\pm 1)\gev$. 
We will discuss the implications that such an assumed signal
would have for the MSSM. While the current statistical significance does
not allow yet to draw firm conclusions on the validity of the above
assumption, our analysis is in fact somewhat more general, as possible
implications of observing (or excluding) a state compatible with a SM-like Higgs
elsewhere in the allowed mass window $115.5\gev<\Mh<127\gev$ \cite{Dec13}
can also be inferred.


\section{MSSM Interpretation}
For calculating the Higgs masses in the MSSM we use the code
{\tt FeynHiggs}~\cite{Degrassi:2002fi,Heinemeyer:1998yj,*Heinemeyer:1998np,*Hahn:2009zz,Frank:2006yh} 
(v. 2.8.5). 
The status of higher-order corrections to the masses (and mixing angles)
in the neutral Higgs sector is quite advanced.%
\footnote{We concentrate here on the case with real parameters. For the
complex case, see~\citeres{Frank:2006yh,Heinemeyer:2007aq} and references
therein.}
The complete one-loop result within the MSSM is available and has been
supplemented by all presumably dominant contributions at the two-loop
level, see \citere{Degrassi:2002fi} for details. Most recently leading
three-loop corrections have been presented~\cite{Martin:2007pg,*Harlander:2008ju}, where the leading term is also included in {\tt FeynHiggs}. 
Following \citere{Degrassi:2002fi}, we estimate 
the (intrinsic) theory uncertainty on the lightest Higgs mass from missing
higher-order corrections to be $\Delta \Mh^{\mathrm{intr}}\sim \pm 2 \gev$. The intrinsic $\Mh$ uncertainties are also somewhat smaller for a SM-like Higgs than in the general case, which makes this estimate conservative.
Concerning the parametric uncertainty from the experimental errors of
the (SM-) input parameters, $\Delta \Mh^{\mathrm{param}}$, the main
effect arises from the experimental error of the top-quark mass. We
incorporate this uncertainty explicitly in our results below by allowing
$\mt$ to vary within the range
$m_t=173.2\pm 0.9\gev$~\cite{Lancaster:2011wr}. Parametric
uncertainties in $\Mh$ from $\als$ are smaller than the $\mt$ uncertainties and will be neglected.
Adding the intrinsic theory uncertainty (conservatively) linearly to the assumed
experimental uncertainty, we arrive at the allowed interval 
\begin{equation}
122\gev< \Mh <128\gev,
\label{eq:mhmssm}
\end{equation}
which will be used for the MSSM interpretation of the assumed Higgs signal.
While for most of this paper we investigate the case where the assumed signal is interpreted as the lighter CP-even Higgs boson, $h$, of the MSSM, 
we comment below also on the possibility of associating the assumed
signal with the {\em second-lightest\/} CP-even Higgs boson, $H$. Since 
the observed excess includes $WW^*$ and $ZZ^*$ final states,
an interpretation in terms of the CP-odd Higgs boson, $A$, appears to be
highly disfavoured.

For our discussions of the possible interpretations of the assumed
signal, we use a phenomenological description of the (CP-conserving) MSSM with
all parameters given at the electroweak scale.  
In order to determine the radiative corrections to the Higgs masses it is
necessary to specify, besides the tree-level parameters $\MA$ and $\tb$,
also the relevant SUSY-breaking parameters entering at higher orders.
In particular,
the parameters in the stop and sbottom sector have a
large 
impact in this context. Since for the case where we interpret the
assumed signal as the lighter CP-even Higgs $h$ we are interested in
particular in
determining lower bounds on the most relevant parameters, we fix those
with smaller impact on $\Mh$ to their values in the \mhmax\ scenario
\cite{Carena:2002qg},  
\begin{equation}
\begin{aligned}
M_1=100\gev,\quad M_2=200\gev\\
 \mgl=0.8\msusy,\quad \mu=200\gev,
\label{eq:M123}
\end{aligned}
\end{equation}
so that conservative lower bounds are obtained for the other parameters.
In \refeq{eq:M123}
$M_{1,2}$ and $\mgl$ are the soft SUSY-breaking gaugino masses
corresponding to the SM gauge group, and $\mu$ is the Higgs mixing parameter.
This choice ensures that the corresponding contributions to $M_h$ are
such that one obtains (approximately) the highest value for $M_h$.
In addition to 
varying the tree-level parameters, we allow for variation in the overall
SUSY mass scale $\msusy$ and the stop mixing parameter 
$\Xt \equiv \At - \mu\CTb$, where $A_{t,b}$ denotes the trilinear coupling
of the Higgs to scalar tops or bottoms. 
We furthermore set $\Ab = \At$. The scalar top masses will be denoted as
$\mste$ and $\mstz$ below, with $\mste \le \mstz$.
It should be noted that when we discuss relatively low values of $\msusy$ 
this refers only to squarks of the third generation (which give rise to
the relevant Higgs mass corrections).
The experimental bounds reported from squark searches
at the LHC \cite{ATLASSusy,*susy11},  on the other hand,
apply only to squarks of the first two generations, which are essentially
irrelevant for Higgs phenomenology. We also do not apply a lower bound on the gluino mass, which leads to more conservative lower limits on the parameters from the Higgs sector  than e.g.~a bound $\mgl>700\gev$~\cite{ATLASSusy,*susy11} would do. We comment further on this point below. As mentioned above,
for the top quark mass we use the latest Tevatron combination
$m_t=173.2\pm 0.9\gev$~\cite{Lancaster:2011wr}, taking the uncertainty
into account by varying $m_t$ over its  $\pm1\,\sigma$ interval.

Besides constraints from the Higgs sector, which we will discuss shortly, one could also consider
indirect constraints on the MSSM parameter space coming from other measurements, such as the anomalous magnetic moment of the muon, \gmt, or from $B$-physics observables such as \bsg.
The former requires in general that $\mu>0$, while the latter is often in better agreement with
experimental data for $\mu\Xt \approx \mu\At < 0$ 
(for a recent analysis see \cite{Mahmoudi:2010xp} and references
therein). We will not apply any indirect constraints here, but 
when presenting the results below we sometimes distinguish between positive
and negative $\Xt$, where the bounds obtained for $\Xt < 0$ could be regarded
as experimentally preferred. However, one should keep in mind that a small
admixture of 
non-minimal flavour violation could bring the \bsg\ results into
agreement with experimental data without changing (notably) the Higgs
sector predictions~\cite{Heinemeyer:2004by,*AranaCatania:2011ak}.


\subsection*{A light CP-even SM-like Higgs boson}
\smallskip
We begin the MSSM interpretation by associating the assumed LHC signal 
with the light CP-even Higgs boson $h$. 
By choosing the relevant parameters such that the 
radiative corrections yield a maximum upward shift to $\Mh$, 
it is possible to obtain lower bounds on the
parameters $\MA$ and $\tb$ governing the tree-level contribution. 
\begin{figure}[t!]
\centering
\includegraphics[width=0.48\columnwidth]{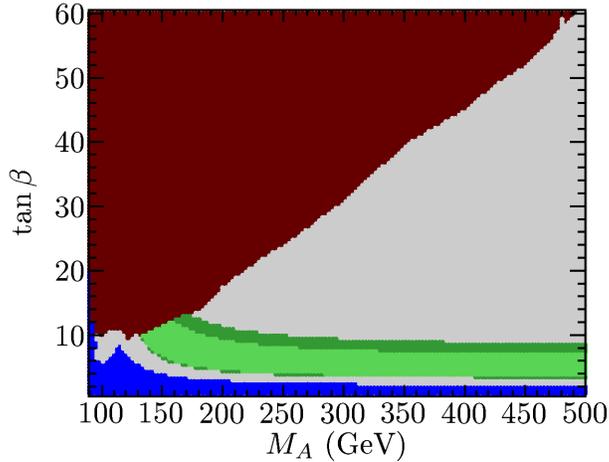}
\caption{Tree-level Higgs sector parameters ($\MA$, $\tb$) for the case
where the parameters governing the higher-order corrections are chosen
such that a maximum value for $\Mh$ is obtained (\mhmax\ benchmark
scenario). The different colours correspond to the regions
  excluded by LEP (blue) and Tevatron/LHC (red). The gray area is the
  allowed parameter space prior to the latest LHC results.  
The green band shows the region where $\Mh$ is compatible with the assumed Higgs
signal (see text).
}
\label{fig:mAtanb}
\end{figure}
The situation where the radiative corrections to $\Mh$ are maximized in
this way is realised in the \mhmax\ scenario
with a stop mixing of $X_t=2\msusy$. In \figref{fig:mAtanb} we show the
result of varying the tree-level parameters in this scenario (with
$\msusy=1\tev$ as originally defined). Constraints on the parameter
space from direct Higgs searches at colliders are taken into account by
using {\tt HiggsBounds} \cite{Bechtle:2008jh,*Bechtle:2011sb}.%
\footnote{We use {\tt HiggsBounds} v.~3.5.0-beta with a private
addition of the latest CMS results on $A/H\to
\tau^+\tau^-$~\cite{CMS-HIG-020}. These new results 
provide the most stringent Tevatron/LHC limits on the ($\MA$, $\tb$)
plane at medium or large $\tb$.} 
~Since we are interpreting an assumed
signal, we do not include the updated exclusion bounds from \cite{Dec13}.
~\figref{fig:mAtanb} shows separately the regions excluded by
LEP \cite{Schael:2006cr} (blue), and the Tevatron/LHC (red). The gray
area is the allowed parameter space before including the bound from
Eq.~\eqref{eq:mhmssm}, and the green band corresponds to the mass interval
compatible with the assumed Higgs signal of
$122\gev<\Mh<128\gev$.
The brighter green is for the
central value for $\mt$, while including also the dark green band
corresponds to a $\pm 1\,\sigma$ variation of $\mt$. 

\begin{table}[b!]
\centering
\begin{tabular}{c|ccc|ccc}
\hline
& \multicolumn{3}{c|}{Limits without $\Mh\sim125\gev$} & \multicolumn{3}{c}{Limits with $\Mh\sim125\gev$}\\
$\msusy$ (GeV) & $\tb$ & $\MA$ (GeV) & $\MHp$ (GeV) & $\tb$& $\MA$ (GeV) & $\MHp$ (GeV)  \\
\hline
500 & $2.7$ & $95$ & $123$ & $4.5$ & $140$ & $161$\\
1000 & $2.2$ & $95$ & $123$ & $3.2$ & $133$ & $155$ \\
2000& $2.0$ & $95$ & $123$ & $2.9$ & $130$ & $152$\\
\hline
\end{tabular}
\caption{Lower limits on the MSSM Higgs sector tree-level parameters
  $\MA$ ($\MHp$) and $\tb$ obtained with and without the assumed Higgs
  signal of $\Mh\sim125\gev$, see \refeq{eq:mhmssm}. The mass limits have been rounded to $1$~GeV.} 
\label{tab:matblimits}
\end{table}
The assumed Higgs signal, interpreted as the lighter CP-even MSSM Higgs
mass, implies in particular that $\Mh>122\gev$ (including theoretical uncertainties), 
which is significantly higher than the limit observed for a SM-like Higgs 
at LEP of $\Mh>114.4$~\cite{Barate:2003sz}. From
\figref{fig:mAtanb} it is therefore possible to extract lower (one parameter)
limits on $\MA$ and $\tb$ from the edges of the green band. 
As explained above, by choosing the parameters entering via
radiative corrections such that those corrections yield a maximum upward
shift to $\Mh$, the lower bounds on $\MA$ and $\tb$ that we have
obtained are general in the sense that they (approximately) hold
for {\em any\/} values of the other parameters.
To address the (small) residual $\msusy$ dependence of the lower bounds on 
$\MA$ and $\tb$, we extract limits
for the three different values $\msusy=\{0.5, 1, 2\}\tev$. The results
are given in Table~\ref{tab:matblimits}, where for comparison we also
show the previous limits derived from the LEP Higgs
searches~\cite{Schael:2006cr}, i.e.\  
before the incorporation of the new LHC results reported in 
\citere{Dec13}.
The bounds on $\MA$ translate directly into lower limits on $\MHp$,
which are also given in the table. A phenomenological consequence of the
bound $\MHp\gtrsim 155\gev$ (for $\msusy=1\tev$) is that it would leave only a
very small kinematic window open for the possibility that MSSM charged Higgs
bosons are produced in the decay of top quarks. 

For deriving the conservative lower bounds on $\MA$ and $\tb$ it
was unnecessary to impose constraints on the production and decay rates
of the assumed Higgs signal in the relevant search channels at the LHC. 
One might wonder whether it would be possible to improve the bound on
$\MA$ by requiring that the rate in the relevant channels should not be
significantly suppressed as compared to the SM case. Such an improvement
would be scenario-dependent, however, i.e.\ the result would depend on
the specific choice made for the other MSSM parameters. We will
therefore not study this issue in further detail.

\begin{figure}[t!]
\centering
\includegraphics[width=0.32\columnwidth]{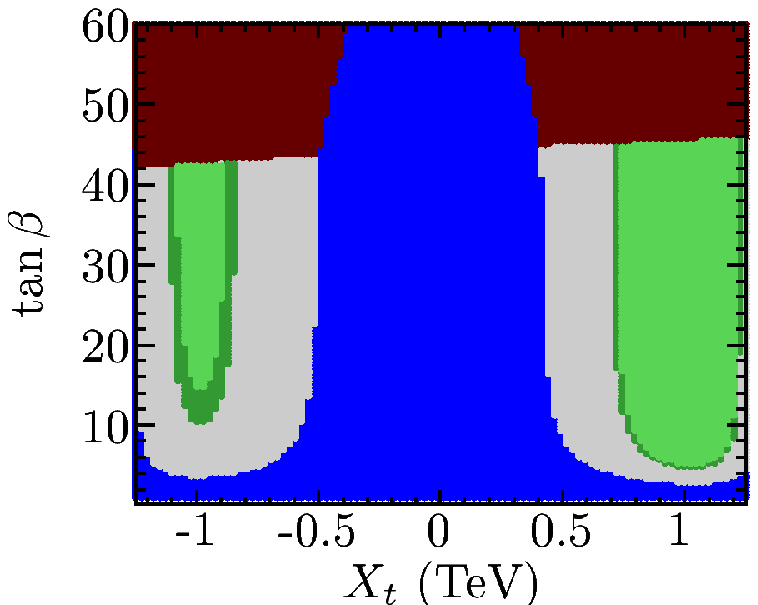}
\includegraphics[width=0.32\columnwidth]{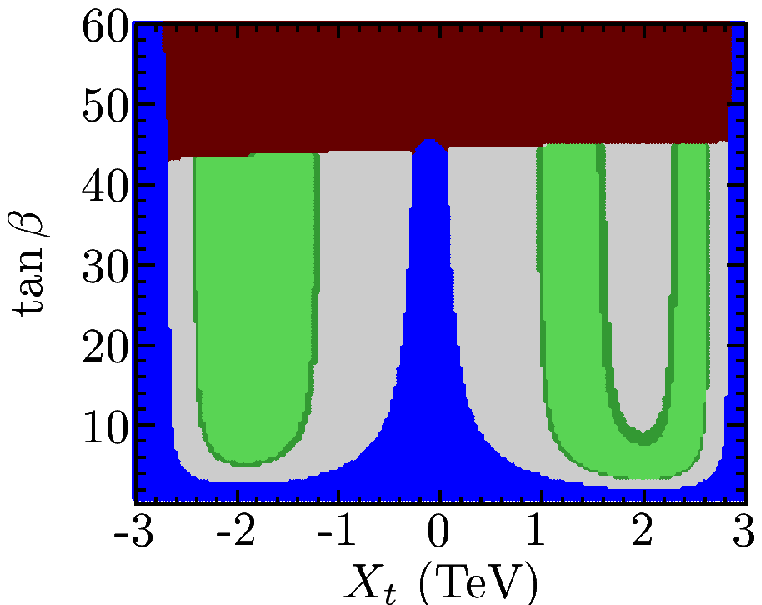}
\includegraphics[width=0.32\columnwidth]{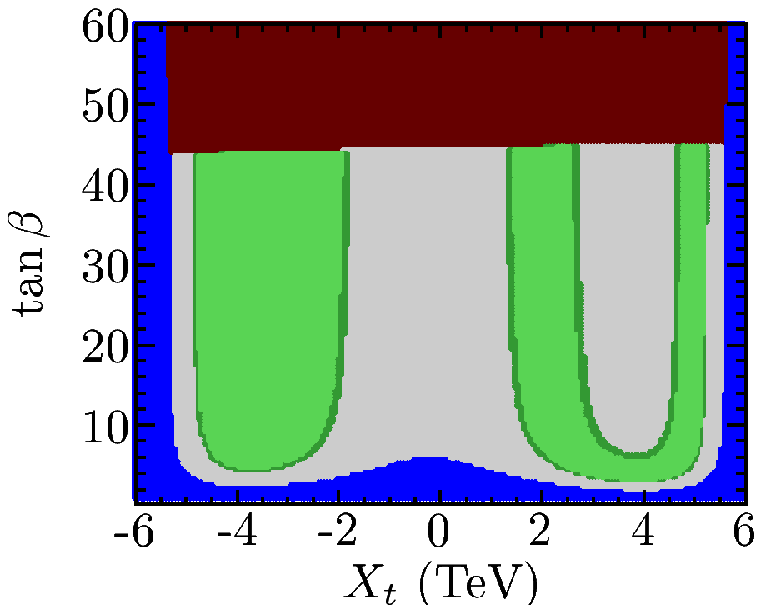}
\caption{Allowed ranges of $\tb$ for $\MA=400\gev$, shown as a function of the stop mixing parameter $X_t$.  The colour coding is as in \figref{fig:mAtanb}. The three plots
correspond to $\msusy=500\gev$ (left), $\msusy=1\tev$ (centre), and $\msusy=2\tev$
(right). } 
\label{fig:xt_tb}
\end{figure}

It might look tempting to extract also an \emph{upper} limit on $\tb$ from 
the green band in \figref{fig:mAtanb}, but in contrast to the lower bound 
which is scenario-independent, this limit will only apply to the specific
case of the \mhmax\ scenario. In fact, the allowed range for $\tb$ depends 
sensitively on the other
parameters, as can be seen from \figref{fig:xt_tb},
where we show the $(\Xt, \tb)$ plane for $\MA = 400 \gev$,
but the results are qualitatively similar for other values of $\MA$  in the
decoupling limit. The main difference is the LHC exclusion limit (in red),
which goes down to lower values of $\tb$ for lower $\MA$. On the other
hand, for $\MA$ in the
non-decoupling regime, even before the new results $\tb$ was already
quite restricted, from above by the the LHC limits, and from below by the LEP
limits, which can also be seen from \figref{fig:mAtanb}. 
The \mhmax\ value of $\Xt = +2\msusy$ turns out to be quite special, since
this parameter region (at least for $\msusy = 1\tev$ and  
$\msusy = 2\tev$) actually shows the highest sensitivity to variations
of $\tb$ when $\Mh\sim125\gev$. This would result in only a narrow allowed
$\tb$ region. For other regions of $X_t$, however, $\tb$ values all the
way up to the LHC bound are compatible with an assumed
signal at $\Mh\sim125\gev$. Further progress could obviously be made if
direct information on the stop sector became available from the LHC or a
future Linear Collider.

\begin{figure}[t!]
\centering
\includegraphics[width=0.48\columnwidth]{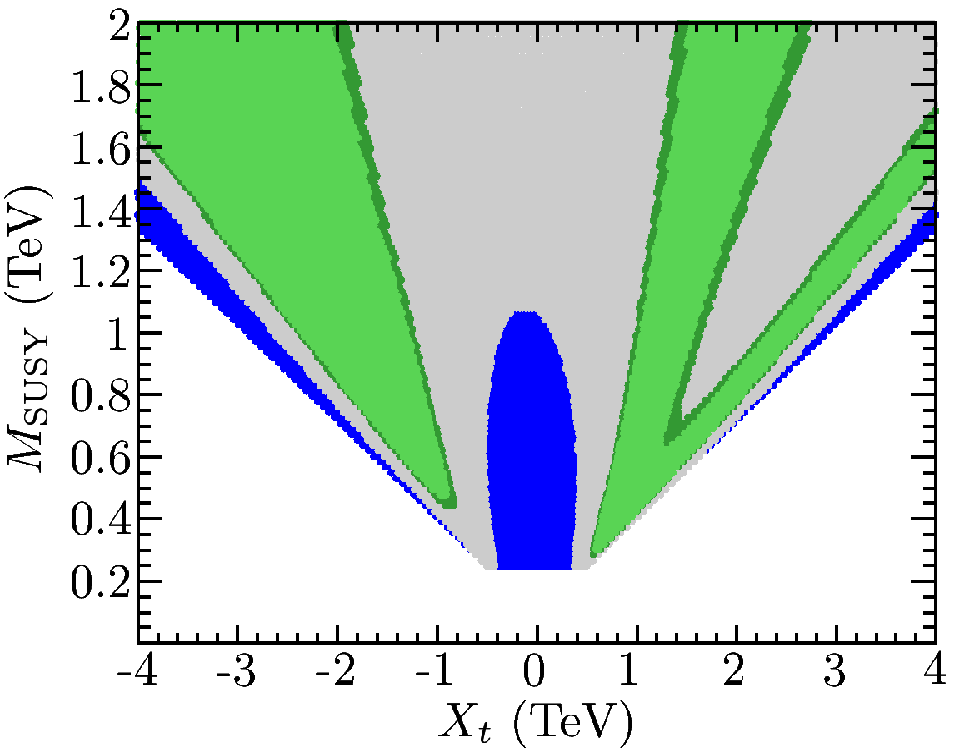}
\includegraphics[width=0.48\columnwidth]{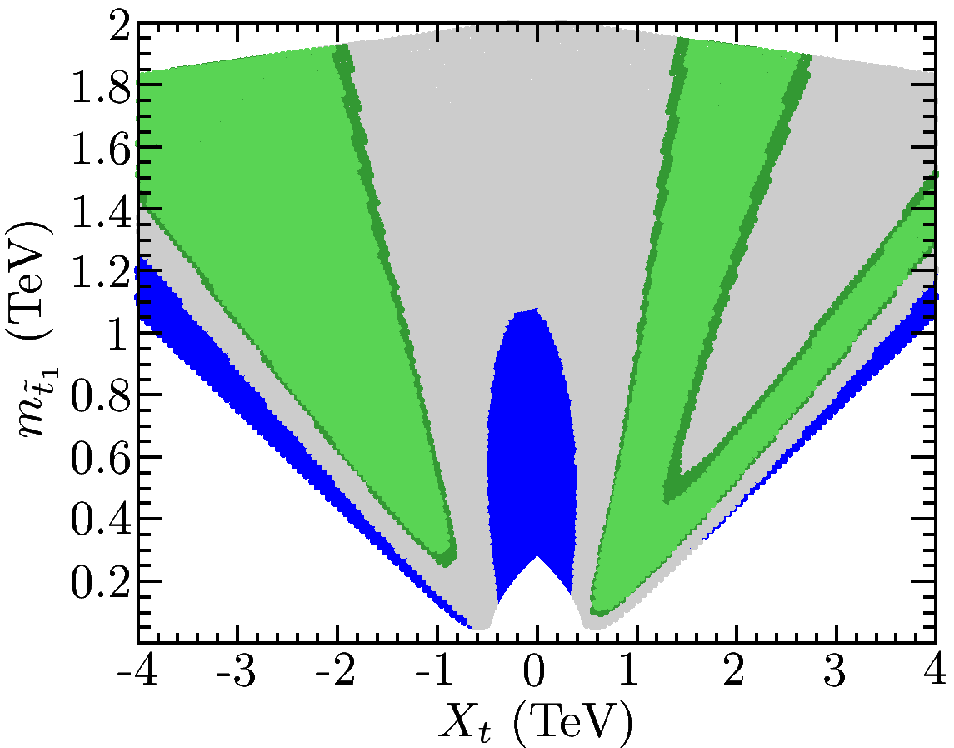}
\caption{Constraints on the MSSM stop sector from the assumed Higgs
signal. 
The allowed ranges are shown in the ($\Xt$, $\msusy$) plane (left) and
the ($\Xt$, $\mste$) plane (right) for
$\MA=1\tev$, $\tb=20$. 
The colour coding is as in \figref{fig:mAtanb}. 
}
\label{fig:xt_msusy}
\end{figure}

Having established lower limits on the tree-level parameters $\MA$
and $\tb$, we now investigate instead what can be inferred from the 
assumed Higgs signal about the higher-order corrections in the Higgs
sector. Similarly to the previous case,
we can obtain an absolute lower limit on the stop mass scale $\msusy$ by
considering the maximal tree-level contribution to $\Mh$. We therefore
perform this analysis in the decoupling limit (fixing $\MA=1\tev$,
$\tb=20$). The resulting constraints for $\msusy$ and $\Xt$ are shown
in~\figref{fig:xt_msusy} (left) using the same colour coding as before.  

Several favoured branches develop in this plane, centred around
$\Xt\sim -1.5\msusy$, $\Xt\sim 1.2\msusy$, and $\Xt\sim 2.5\msusy$. 
The minimal allowed
stop mass scale is $\msusy\sim 300\gev$ with positive $\Xt$ and
$\msusy\sim 500\gev$ for negative $\Xt$ (which is in general
preferred by \bsg, see above).
The results on the stop sector can also be interpreted as a lower limit
on the mass $\mste$ of the lightest stop squark. This is shown
in \figref{fig:xt_msusy} (right). It is interesting to note from the
figure that without the assumed Higgs signal, there is essentially no
lower bound on the lightest stop mass coming from the Higgs
sector. Taking the new results into account, we obtain the lower bounds
$\mste>100\gev$ ($\Xt>0$) and $\mste>250\gev$ ($\Xt<0$). 
These bounds can be compared to those from direct searches, where the
LEP limit $\mste \gsim 95 \gev$ is still valid \cite{Nakamura:2010zzi}. Results from stop searches at the Tevatron can also be found in this reference. No new stop limits have been established so far from the SUSY searches at the LHC~\cite{ATLASSusy,*susy11}.
It should be noted that our stop mass bound is rather conservative, since the low mass scales discussed here correspond to a gluino mass $\mgl=0.8\,\msusy < 300\gev$, which is  experimentally disfavoured \cite{ATLASSusy,*susy11,Nakamura:2010zzi,ATLAS-CONF-2011-030}. Since the low gluino mass contributes towards a higher value of $\Mh$, a lower bound on $\mgl$ would lead to a stronger bound on $\mste$. As an example, in a simplified model consisting just of the gluino, the
squarks of the first two generations and a massless lightest
supersymmetric particle, the ATLAS Collaboration has inferred a lower
bound of about $700\gev$ on $\mgl$~\cite{ATLASSusy}. Imposing such a bound
on $\mgl$ in our analysis would shift the lower limit on $\mste$ to $\mste\gtrsim 200\gev$ ($\mste\gtrsim 350\gev$) for positive (negative) $X_t$. It should be noted, however, that in the presence of a light stop decays
of the gluino into a top and a scalar top would open up, $\tilde g \to \tilde t_1 t$, which are expected to weaken the bound on
$\mgl$ as compared to the analysis in the simplified model where this
decay mode is assumed to be absent.


\subsection*{A heavy CP-even SM-like Higgs boson}
\smallskip
All results presented up until this point apply only if we interpret the
assumed signal as corresponding to the light CP-even MSSM Higgs $h$. We
now discuss briefly the alternative possibility that the heavier
CP-even $H$ has a mass $\MH\sim 125\gev$ (with the same experimental and
theoretical uncertainties as before, see \refeq{eq:mhmssm}) and SM-like properties.

\begin{figure}[t!]
\centering
\includegraphics[width=0.48\columnwidth]{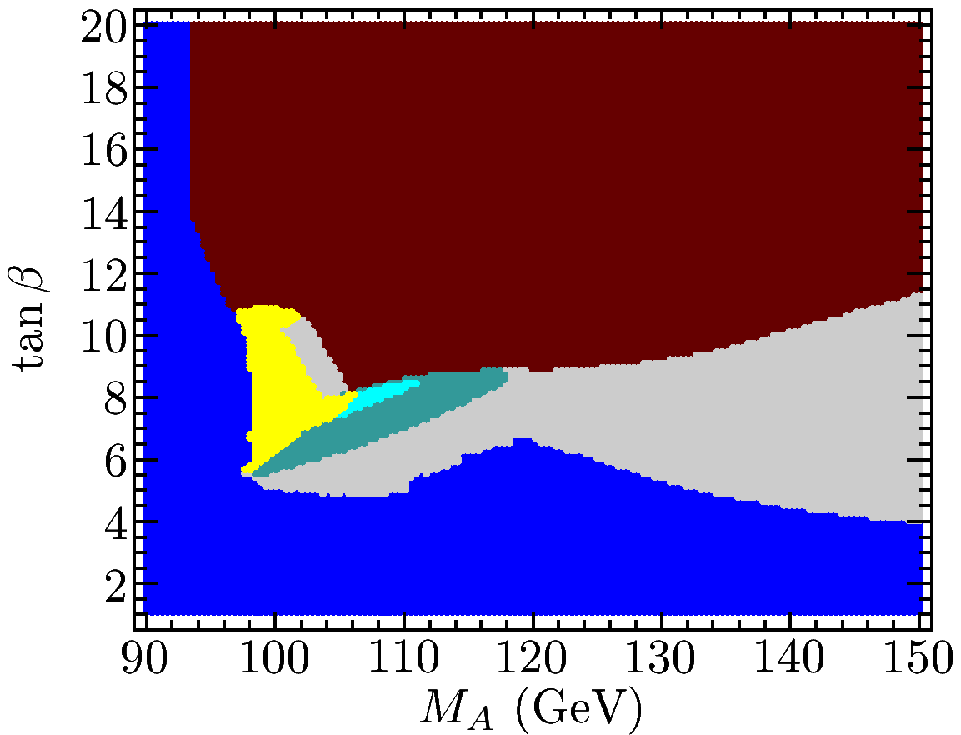}
\includegraphics[width=0.48\columnwidth]{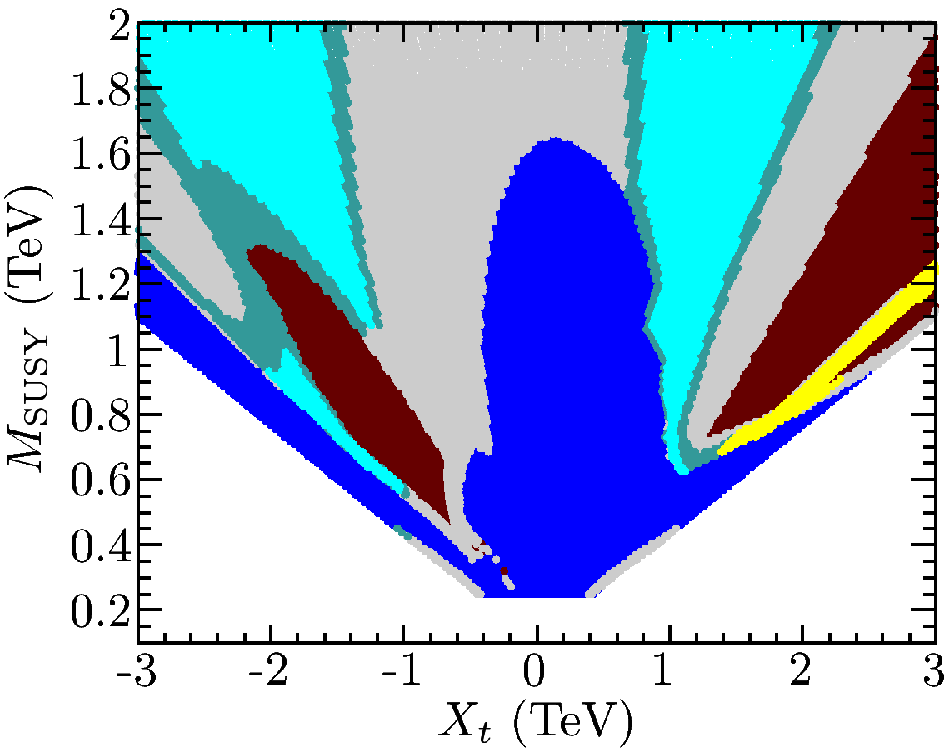}
\caption{Parameter space in the alternative $\MH\sim125\gev$
  scenario. The colour coding is similar to \figref{fig:mAtanb}, with
  new regions (cyan and yellow) where $\MH$ is in the range compatible with the assumed $H$ signal. In addition, for the yellow region the heavy Higgs has a rate for
production times decay into $\gamma\gamma$ of at least $90\%$ of the corresponding SM values. For the plot in the $(\MA, \tb)$ plane (left) we have assumed $\msusy=1\tev$, $\Xt=2.3\tev$
  and for the stop parameters (right) we fix $\MA=100\gev$, $\tb=10$.
In both cases $\mu=1\tev$, and the remaining parameters are given by Eq.~\eqref{eq:M123} with the additional requirement $\mgl>700\gev$.} 
\label{fig:Hscenario}
\end{figure}

In order to investigate whether there is a region in the MSSM parameter
space that admits this
solution we performed a scan over the relevant free parameters ($\MA$,
$\tb$,
$\msusy$, $\Xt$), keeping $\mu=1\tev$ fixed and the remaining parameters according to Eq.\eqref{eq:M123}.
The results are shown in \figref{fig:Hscenario},
indicating the region where $\MH$ fulfills Eq.~\eqref{eq:mhmssm} by cyan
colour to distinguish it from the case discussed above (similarly to above, the darker region corresponds to the variation of $m_t$).
As we can see from this figure, it is possible to obtain $\MH$ in
the right range in a region with low $\MA$ and moderate $\tb$ (left plot) where we have set $\msusy=1\tev$, $\Xt=2.3\tev$.  In the right plot we set $\MA=100\gev$, $\tb=10$ and show the regions compatible with a
heavier CP-even Higgs having a mass $\MH\sim125\gev$ in the plane of the
stop sector parameters $\msusy$ and $X_t$. We find that such an
interpretation is possible over extended regions of the ($\msusy$, $X_t$)
parameter plane. Requiring in addition that the production and
decay rates into $\gamma\gamma$ and vector bosons are at least 90\% of
the corresponding SM rates, a smaller allowed region is found (yellow) 
with large values for the stop mixing ($X_t\gtrsim 1.5\tev$). In the
yellow region enhancements of the rate of up to a factor of three as
compared to the SM rate are possible.
Concerning the mass of the lighter CP-even Higgs boson $h$ in this kind
of scenario we we find in our scan allowed values for $\Mh$ 
only {\em below\/} the SM LEP limit of
$114.4\gev$~\cite{Barate:2003sz} (with reduced couplings to gauge bosons
so that the limits from the LEP searches for non-SM like Higgs bosons
are respected~\cite{Schael:2006cr}). A particularly intriguing option could be
$\MH \simeq 125\gev$, $\Mh \simeq 98 \gev$, in view of the fact that LEP
observed a certain excess at $\Mh \simeq 98 \gev$ \cite{Schael:2006cr}
(whose interpretation is of course subject to the look--elsewhere
effect). This combination of Higgs masses is realized (with $H$ SM-like),
for instance, for $\msusy=1\tev$, $\Xt=2.4\tev$, $\mu=1\tev$, $\MA=106\gev$,  and
$\tb=7$. 
For this scenario we find a reduced coupling $(g_{hZZ}/g_{HZZ}^{\SM})^2 = 0.1$ of the lightest Higgs boson to a pair of
$Z$~bosons.

Despite the available parameter space, it should be noted that the scenario 
where the heavier CP-even Higgs is SM-like and has a mass of 
$\MH\sim 125\gev$ appears somewhat more contrived than the $h$ interpretation. 
In particular, we find that simultaneously large values for the
$\mu$ parameter and a large mixing in the stop sector are required in
order to obtain a SM-like rate of production and decay of the heavy
CP-even Higgs in the relevant channels. We leave a more detailed
investigation of this scenario for future work.

 
\section{Conclusions}
An excess in the SM-like Higgs searches at ATLAS and CMS has recently
been reported~\cite{Dec13} around $\MHSM\simeq 125\gev$, which within the experimental uncertainties appears to be remarkably consistent between ATLAS and CMS and is supported by
several search channels. While it would be premature to assign more
significance to this result than regarding it as a possible
(exciting) hint at this
stage, it is certainly very interesting to note that this excess has
appeared precisely in the region favoured by the global fit within the
SM, and within the range predicted in the MSSM. Concerning the MSSM, it
is remarkable that the mass region above the upper MSSM bound on a light
SM-like Higgs is meanwhile ruled out~\cite{Dec13}. Observing a state
compatible with a SM-like Higgs boson with
$\MHSM> 135\gev$ would have unambiguously ruled out the MSSM (but would have
been viable in the SM and in non-minimal supersymmetric extensions of
it). We therefore regard the reported results as a strong motivation for 
studying the possible interpretation of an assumed (still hypothetical,
of course) signal at $125\gev \pm 1\gev$. In this paper we have discussed the
possible implications of such an assumed signal within the MSSM, where 
we have investigated both the possibilities that the assumed signal is associated with the light CP-even Higgs boson of the MSSM, $h$, and the (slightly more exotic) possibility that the assumed signal in fact corresponds to the heavier CP-even Higgs boson
$H$.

Investigating the interpretation $\Mh = 125 \pm 1 \gev$ first, we have
demonstrated that there is a significant parameter space of the MSSM
compatible with the interpretation that the assumed signal corresponds
to the lighter CP-even MSSM Higgs boson. While it would not be
appropriate to assign any physical significance to point densities in
MSSM parameter space, our scans nevertheless do not seem to indicate a strong case for going from the MSSM to non-minimal
SUSY models even though the reported excess is not very far away from
the upper bound on the lightest Higgs mass in the MSSM. It should be
noted that the question to what extent the scenarios discussed in this paper can be realized in constrained GUT-based models
of SUSY breaking is of a very different nature. We do not pursue
this any further here, besides mentioning that it has already been shown 
to be rather difficult to get to such high $\Mh$
values in models such as the CMSSM, mGMSB, mAMSB, or NUHM1~\cite{Heinemeyer:2008fb,*Carena:2010ev,*Buchmueller:2011ki,*Buchmueller:2011sw}. 

We performed two kinds of complementary investigations of the
implications of an assumed Higgs signal at
$\Mh = 125 \pm 1 \gev$. Setting the parameters that enter via the 
(in general) numerically large higher-order corrections in the MSSM
Higgs sector to their values in the \mhmax\ benchmark
scenario, which maximizes the upward shift in $\Mh$ as compared to the
tree-level value, we have obtained conservative lower limits on the
parameters governing the $\Mh$ prediction at tree level, $\MA$ and
$\tb$. We have found that an assumed signal of $\Mh = 125 \pm 1 \gev$
(when including conservatively estimated intrinsic theoretical uncertainties from  
unknown higher orders, and taking into account the most important
parametric uncertainties arising from the experimental error on the
top-quark mass) yields the lower bounds $\MA>133\gev$ and $\tb>3.2$ (for
$\msusy=1\tev$). The bound on $\MA$ translates directly into a lower limit $\MHp > 155\gev$, which restricts the kinematic window for MSSM charged Higgs production in the decay of top quarks.

Choosing values for $\MA$ and $\tb$ in the decoupling region, in a
second step we have investigated the constraints on the scalar top and
bottom sector of the MSSM from an assumed signal at $\Mh = 125 \pm 1 \gev$.
In particular, we have found that a lightest stop
mass as light as $\mste \sim 100\gev$ is still compatible with the
assumed Higgs signal. The bound on $\mste$ raises to $\mste\gsim
250\gev$ if one restricts to the negative sign of the stop mixing
parameter $\Xt \equiv \At - \mu/\tb$, which in general yields better
compatibility with the constraints from ${\rm BR}(b \to s \gamma)$.

As an alternative possibility, we have investigated in how far it is
possible to associate the assumed Higgs signal with the heavier CP-even
Higgs boson $H$. Performing a scan over $\MA$, $\tb$, $\msusy$ and $\Xt$
we have found an allowed area at low $\MA$ and moderate $\tb$.
A SM-like rate for production and decay of the heavier CP-even
Higgs in the relevant search channels at the LHC is possible 
for large values of $\mu$ and large mixing in the stop sector.
It is interesting to note that in the scenario where the assumed
Higgs signal is interpreted in terms of the heavier CP-even Higgs boson
$H$ the mass of the lighter Higgs, $\Mh$,
always comes out to be {\em below\/} the SM LEP limit
of $114.4\gev$ (with reduced couplings to gauge bosons
so that the limits from the LEP searches for non-SM like Higgs bosons
are respected). The fact that scenarios like this are in principle
viable should serve as a strong motivation for extending the LHC Higgs
searches, most notably in the $\ga\ga$ final states, also to the mass
region below $100\gev$. 

Needless to say, an MSSM interpretation of the observed excess would of
course gain additional momentum if the searches for the scalar quarks of
the third generation and the direct searches for the colour-neutral
SUSY states, which so far have resulted in only very weak limits, would
soon give rise to a tantalising excess (or more than one) as well.


\section*{Acknowledgments}
We thank Johan Rathsman and Rikard Enberg for useful suggestions at an
early stage of this project. We also thank Tim Stefaniak and Oliver
Brein for discussions and help with {\tt HiggsBounds}, in particular on
the CMS $A\to \tau^+\tau^-$ results.  
We thank Paloma Arenas Guerrero for her contributions to our
investigation of a possible heavy CP-even SM-like Higgs boson. 
This work has been supported by the Collaborative Research Center SFB676
of the DFG, ``Particles, Strings, and the Early Universe''. 
The work of S.H.\ was supported in part by CICYT (grant FPA 2010--22163-C02-01) and by the Spanish MICINN's Consolider-Ingenio 2010 Program under grant MultiDark CSD2009-00064. 

\bibliographystyle{JHEP}
\bibliography{higgs}

\end{document}